\documentclass[aps,preprintnumbers,superscriptaddress,showpacs,fleqn,twocolumn,longbibliography]{revtex4}

\usepackage{booktabs}
\usepackage{epsfig}
\usepackage{psfrag}
\usepackage{amsfonts}
\usepackage{amsmath}
\usepackage{graphicx}
\usepackage{float}
\usepackage{dcolumn}
\usepackage{bm}

\begin{document}

\title{A study on the exotic state ${P_{c}(4312),P_{c}(4440),P_{c}(4457)}$ in $pp$ collisions at $\sqrt{s}=7$, 13~TeV}

\author{Chun-hui Chen$^{1}$, Yi-Long Xie$^{1}$, Hong-ge Xu$^{1}$, Zhen Zhang$^2$, Dai-Mei Zhou$^2$, Zhi-Lei She$^1$, Gang Chen$^{}$\footnote{Corresponding Author: chengang1@cug.edu.cn}   }

\address{
${^1}$School of Mathematics and Physics, China University of Geoscience, Wuhan 430074, China\\
${^2}$Institute of Particle Physics, Huazhong Normal University, Wuhan 430082, China}

\begin{abstract}
The exotic resonant state $P_{c}^{\pm}(4312)$, $P_{c}^{\pm}(4440)$, and $P_{c}^{\pm}(4457)$ are studied by using the dynamically constrained phase space coalescence({\footnotesize DCPC}) model and {\footnotesize PACIAE} model in $pp$ collisions at $\sqrt{s}=7$, 13~TeV, respectively. We consider exotic state $P_{c}^{\pm}(4312)$, $P_{c}^{\pm}(4440)$ and $P_{c}^{\pm}(4457)$ to be three kinds of possible structures, i.e. pentaquark state, the nucleus-like state, and the molecular state based on $P_c^\pm \to J/\psi{p}{(\bar p)}$ bound state. The yield, the transverse momentum distribution and the rapidity distribution of $P_{c}$ with three different structures are all predicted. The results indicated that the yield is on the order of $10^{-6}$, which might provide reference for experimental research.

\end{abstract}

\maketitle
\section{Introduction}
Quantum chromodynamics (QCD) as the fundamental theory of the strong interactions, specifies that quarks are bound in color-neutral hadrons with various structures. In the conventional quark model, a meson is made up a quark-antiquark pair and a baryon is made of three quarks. However, there exist some other interpretations for baryon structures which allows more complex structure such as multiquark states~\cite{Ref1}, hadronic moleculesons~\cite{Ref2} hybrid states~\cite{Ref3,Ref4} and glueball~\cite{Ref60}. These complex structures are called exotic state hadrons. Exotic hadrons will continuously be a central topic in hadron physics.

In the last few years, various exotic state hadrons are observed in high energy collision experiments~\cite{Ref5,Ref6}, are called the XYZ particles. Particularly in 2015, LHCb Collaboration~\cite{Ref7} reported two hidden-charm pentaquarks, i.e. $P_{c}^{+}(4450)$ and $P_{c}^{+}(4380)$ in the $J/\psi{p}$ invariant mass distribution of the decaying $\Lambda_{B}\to K^{-}J/\psi{p}$. $P_{c}^+(4450)$ showed up a narrow peak in distribution with a width of $39\pm5\pm19$ MeV and $P_{c}^{+}(4380)$ showed up another peak with a width of $205\pm18\pm86$ MeV. In 2019~\cite{Ref8}, LHCb Collaboration updated their measurement and observed the narrow peak $P_{c}(4450)$ were separated into two narrow overlapping peaks, $P_{c}^{+}(4440)$ and $P_{c}^{+}(4457)$, with a width of $20.6\pm{4.9}_{-10.1}^{+8.7}$ and $6.4\pm{2.0}_{-1.9}^{+5.7}$, respectively. Meanwhile, the third peak $P_{c}^{+}(4312)$ shows up and $P_{c}^{+}(4380)$ are too small to be observed.

$P_{c}$ states have minimal quark content $duuc\overline{c}$. For the two hidden-charm pentaquarks, there are different theoretical interpretations: the loosely bound meson-baryon molecular states~\cite{Ref9,Ref10,Ref11}, the tightly bound pentaquark states~\cite{Ref12}, and the hadrocharmonium states~\cite{Ref13}. The $P_{c}$ states with narrow peaks are under the threshold of binding energies of $\Sigma_{c}^{+}\overline{D}^{0}$, $\Sigma_{c}^{+}\overline{D}^{*0}$, which makes the bound states of baryon and meson possible~\cite{Ref8}. Hence we suggest that the $P_{c}$ states could be interpreted as the bound states of $J/\psi{p}{(\bar p)}$~\cite{Ref14}.  Similar to Ref.~\cite{Ref15} we also treated $P_{c}(4312)$, $P_{c}(4440)$ and $P_{c}(4457)$ as three structures, i.e. pentaquark state, nucleus-like state and molecular state, according to the different distances of corresponding components.

In this paper, we study $P_{c}^{\pm}$ states through decaying $P_{c}^{\pm}\to J/\psi {p}{(\bar p)}$ by using Monte Carlo simulation~\cite{Ref16} approach in $pp$ collisions at $\sqrt{s}=7$, 13~TeV. As the first step, the parton and hadron cascade model ({\footnotesize PACIAE})~\cite{Ref17} is used to generate the hadronic final state, including $J/\psi$ and $p(\bar p)$; then the $P_{c}(4312)$, $P_{c}(4440)$ and $P_{c}(4457)$ states yield, as well as the transverse momentum distribution and the rapidity distribution are predicted by a dynamically constrained phase-space coalescence({\footnotesize DCPC}) model~\cite{Ref18}. Here we will study three different configurations of the $P_c$ state namely pentaquark state, nucleus-like state and molecular state.

\section{{\footnotesize PACIAE} Model and {\footnotesize DCPC} Model}

The parton and hadron cascade model {\footnotesize PACIAE}~\cite{Ref17} is based on PYTHIA 6.4~\cite{Ref16} to describing various relativistic collision modes, such as relativistic $pp$, $\overline{p}p$, $ \overline{e}e$ collisions, and relativistic nuclear collisions (A-A and p-A). The {\footnotesize PACIAE} model is composed of four main stages: parton initiation, parton rescattering, hadronization, and hadron rescattering.

In order to produce the parton initial state, the string fragmentation~\cite{Ref17} is temporarily turned off in {\footnotesize PACIAE} and breaks up the diquarks (anti-diquarks) to produce some mater composed by quarks anti-quarks and gluon. The parton rescattering is followed which depend on the  $2\to 2$ LO-pQCD parton-parton cross sections~\cite{Ref19}. After all parton rescatterings, hadronization of the final state partons are computed by the string fragmentation~\cite{Ref17} or the Monte Carlo coalescence model~\cite{Ref16}. The last stage is hadron rescattering process happening among two-body collision until the hadronic freeze-out.

In this paper, the yields of bound states are usually calculated in two steps. First, the final state hadrons are calculated by the {\footnotesize PACIAE} model. Then, the bound states or exotic states are combined by the {\footnotesize DCPC} model\cite{Ref18}. The {\footnotesize DCPC} model is developed to calculate production of the bound states or exotic states after the final state particles produced by the {\footnotesize PACIAE} model~\cite{Ref17} in $pp$ collisions.

From quantum statistical mechanics~\cite{reff1}, one can not precisely define both position $\vec q\equiv (x,y,z)$ and momentum $\vec p \equiv (p_x,p_y,p_z)$ of a particle in six-dimensional phase space because of the uncertainty principle, $\Delta \vec q\Delta \vec p \sim h^3$. One can only say this particle lies somewhere within a six-dimensional quantum box or state of volume of $\Delta \vec q\Delta \vec p$ volume element in the six-dimensional phase space corresponds to a state of the particle. So we can estimate the yield of a single particle~\cite{reff1} by
\begin{equation}
Y_1=\int_{{E}_{A}\leq{H}\leq{E}_{B}} \frac{d\vec qd\vec p}{h^3},
\end{equation}
where $H$ and ${E}_{A}$, ${E}_{B}$ represent the energy function (Hamiltonian) and the energy threshold of the particle, respectively. The variables $\vec{p}$ and $\vec{q}$  are the coordinates and momentum of the particle in the center-of-mass frame at the moment after the hadronic freeze-out. Furthermore, the yield of $N$ particles clusters can be calculated by the following integral:
\begin{equation}
Y_{N} =\int\cdots\int_{{E}_{A}\leq{H}\leq{E}_{B}}{\frac{d\vec q_{1}d\vec p_{1}\cdots d\vec q_{N}d\vec p_{N}}{(h)^{3N}}}.
\end{equation}

Therefore, the yield of a  $P_{c}(4312)$, $P_{c}(4440)$ and $P_{c}(4457)$ consisting of $J/\psi{p}{(\bar p)}$ cluster in the {\footnotesize DCPC} model can be calculated by
\begin{align}
            &Y_{P_{c}^{\pm}\rightarrow J/\psi {p}{(\bar p)}} =\int ...\int\delta_{12}\frac{d\vec q_{{p}{(\bar p)}}d\vec p_{{p}{(\bar p)}} d\vec q_{J/\psi} d\vec p_{J/\psi}}{h^{6}},
\label{yield} \\
&\delta_{12}=\left\{
  \begin{array}{ll}
  1 \hspace{0.2cm} \textrm{if} \hspace{0.2cm} 1\equiv {p}{(\bar p)}, 2\equiv J/\psi;\\
    \hspace{0.2cm} ~m_0-\Delta m\leq m_{inv}\leq m_0+\Delta m;\\
      ~~~q_{12}\leq R_{0};\\
  0 \hspace{0.2cm}\textrm{otherwise}.
  \end{array}
  \right.
\label{yield1}
\end{align}
\textrm{where},
\begin{equation}
\hspace{0.5cm}  m_{inv}=\sqrt{(E_{p(\bar p)}+E_{J/\psi})^2-(\vec p_{p(\bar p)}+\vec p_{J/\psi})^2}.
\label{yield2}
\end{equation}

Where $m_0$ denotes the rest mass of $P_{c}$ states, $\Delta m$ refers to its mass uncertainty, $q_{12}= |\vec q_1-\vec q_2|$ represents the distance between the two particles $p$ or ${\bar p}$ and $J/\psi$. The dynamic constraint ${E}_{A}\leq{H}\leq{E}_{B}$ in Eq.1 is replaced by the $m_0-\Delta m\leq m_{inv}\leq m_0+\Delta m$. The rest masses of $P_{c}(4312)$, $P_{c}(4440)$ and $P_{c}(4457)$  were $m_0=4311.9, 4440.3, 4457.3$ MeV/c$^2$~\cite{Ref14}, respectively.

$P_{c}(4312)$, $P_{c}(4440)$ and $P_{c}(4457)$  is constructed by the combination of hadrons $p (\bar p)$ and $J/\psi$ after the final hadrons have been produced by {\footnotesize PACIAE} model in $pp$ collisions at $\sqrt{s}=7$, 13~TeV. According to the different distances $q_{12}$ between $p (\bar p)$ and $J/\psi$, the exotic state $P_c$ can be divided three structures: the pentaquark refers $R_0<0.9$ fm, the nuclear-like state refers $0.9\le R_0<1.91$ fm, and the molecular state refers $R_0\ge1.91$ fm~\cite{Ref14,Ref21}.

\section{Results}

First, the multiparticle final state are produces using the {\footnotesize PACIAE} model in $pp$ collisions at $\sqrt{s}=7$, 13~TeV. In addition to the $K$ factor and parameters parj(1), parj(2), and parj(3), the other model parameters are fixed on the default values given in the PYTHIA model. In this model, parj(1) is the suppression of diquark-antidiquark pair production compared with quark-antiquark production, parj(2) is the suppression of strange quark pair production compared with up or down pair production, and parj(3) is the extra suppression of strange diquark production compared with the normal suppression of strange quark~\cite{Ref17}.
We determine the $K$ factor, parj(1,2,3) for primary hadrons in {\footnotesize PACIAE} model
by fitting to the LHCb and ALICE data of $J/\psi$, $p$ and $\bar p$~\cite{Ref22,Ref23}. The fitted values of $K$ = 0.9, parj(1) = 0.13, parj(2) = 0.20, and parj(3) = 0.90 for $pp$ collisions are used in later calculations.

Table I shows the yield of $p (\bar p)$ and $J/\psi$ calculated with $|y|<0.5$, $0.6<p_T<6$ GeV/c for ${p}{(\bar p)}$ and $2.0<y<4.5$, $0<p_T<14$ GeV/c for $J/\psi$, separately. Obviously the results of the {\footnotesize PACIAE} model are consistent with the data measured by LHCb Collaboration and ALICE Collaboration~\cite{Ref22,Ref23}.

\begin{table}[hp]
\caption{~The yield of particles ($p$ or $\bar p$ and $J/\psi$) in $pp$ collisions at $\sqrt{s}=7$ TeV simulated by the {\footnotesize PACIAE} model, and compared with data from LHCb Collaboration and ALICE Collaboration as the $|y|<0.5$, $0.6<p_T<6$ GeV/c for $p (\bar p)$ and $2.0<y<4.5$, $0<p_T<14$ GeV/c for $J/\psi$, respectively~\cite{Ref22,Ref23}. Here, $J/\psi$ is from b decay and prompt production.}	
\centering
\renewcommand{\arraystretch}{1.2}
\begin{tabular}{ccc} \hline  \hline
particles &        Experiment data       & {\footnotesize PACIAE}\\ \hline
$J/\psi$  & $(1.64\pm0.007\pm0.22) \times 10^{-4}$ & $(1.680\pm0.002)\times10^{-4}$\\
$p$     & $0.124\pm0.009 $                         & $0.123\pm0.001$ \\
$\bar p$     & $0.123\pm0.010$                          & $0.122\pm0.001$\\ \hline \hline
\end{tabular} \label{paci1}
\end{table}

\begin{figure*}[t]
\includegraphics[width=0.85\textwidth,height=0.55\textwidth]{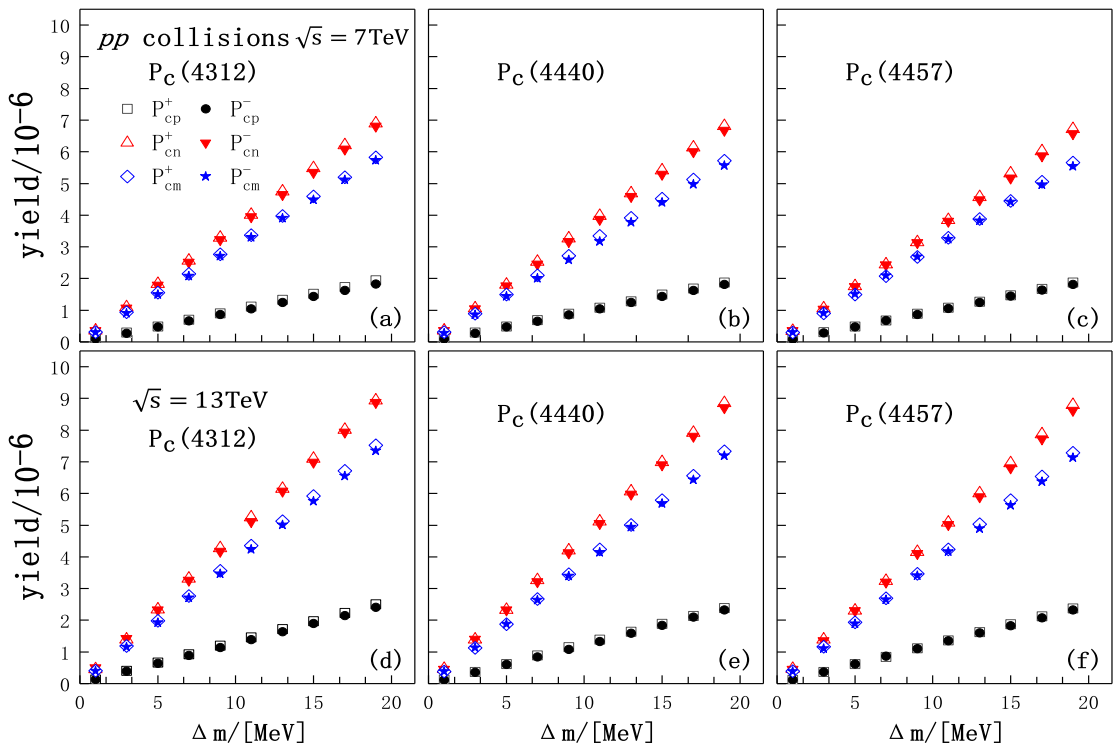}
\centering
\caption{The distribution of the yield of exotic states $P_{c}^{\pm}(4312)$, $P_{c}^{\pm}(4440)$, and $P_{c}^{\pm}(4457)$ with three structures ($P_{cp}, P_{cn}, P_{cm}$) in $pp$ collisions at $\sqrt{s}=7$, 13~TeV, as a function of mass uncertainty $\Delta m$. The data are computed by using {\footnotesize PACIAE+DCPC} model based on the $J/\psi{p}{(\bar p)} $ bound state. (a) as exotic states $P_{c}^{\pm}(4312)$ at $\sqrt{s}=7$~TeV, (b) as exotic states $P_{c}^{\pm}(4440)$ at $\sqrt{s}=7$~TeV, (c) as exotic states  $P_{c}^{\pm}(4457)$ at $\sqrt{s}=7$~TeV, (d) as exotic states $P_{c}^{\pm}(4312)$ at $\sqrt{s}=13$~TeV, (e) as exotic states $P_{c}^{\pm}(4440)$ at $\sqrt{s}=13$~TeV, (f) as exotic states  $P_{c}^{\pm}(4457)$ at $\sqrt{s}=13$~TeV.}
\label{tu1}
\end{figure*}

Then, we generate the event final states including $p (\bar p)$ and $J/\psi$ using the {\footnotesize PACIAE} model in $pp$ collisions at $\sqrt{s}=7$, 13~TeV, with the $|y|<3$, $0.<p_T<10$ GeV/c. Inputting $J/\psi$ and $p{(\bar p)}$ into the {\footnotesize DCPC} model, the exotic states $P_c$ are combined to generate by constructing $J/\psi p{(\bar p)}$ clusters, which can be considered as the exotic states $P_{c}(4312)$, $P_{c}(4440)$ and $P_{c}(4457)$. Here, $J/\psi$ is from all $J/\psi$ sources. We assume that there are three structures of $P_c$ states~\cite{Ref15}: pentaquark state, nucleus-like state and molecular state, which are denoted as $P_{cp}$, $P_{cn}$ and $P_{cm}$, respectively.

\begin{large}
\small\addtolength{\tabcolsep}{2.6pt}
\begin{table*}[tbp]
\caption{The yields $10^{-6}$ of exotic states $P_{c}^{\pm}(4312)$, $P_{c}^{\pm}(4440)$ and $P_{c}^{\pm}(4457)$ with three structures computed by {\footnotesize PACIAE+DCPC} model in $pp$ collisions at $\sqrt{s}=7$, 13~TeV.
Their parameter values $\Delta m$ is taken as $\Delta m=\Gamma/2= 4.9$ MeV for $P_{c}^{\pm}(4312)$, as 10.3 MeV for $P_{c}^{\pm}(4440)$, and as 3.2 MeV for $P_{c}^{\pm}(4457)$. The value of decay width $\Gamma$ is taken from LHCb experiment~\cite{Ref8}, respectively.}
\renewcommand{\arraystretch}{1.2}\setlength{\tabcolsep}{3.mm}
\begin{tabular}{ccccccc} \hline\hline
$\sqrt{s}$& bound state & Particle & Pentaqaurk ($P_{cp}$)& Nucleus-like ($P_{cn}$)& Molecular ($P_{cm}$)& total \\ \hline
   &  & $P_c^+(4312)$ & $0.49\pm0.01$ & $1.79\pm0.04$ & $1.51\pm0.05$ & $3.79\pm0.10$ \\
  & $J/\psi p$ & $P_{c}^+(4440)$ &$1.02\pm0.04$ & $3.71\pm0.10$ & $3.63\pm0.09$ & $8.34\pm0.17$ \\
7TeV &   & $P_{c}^+(4457)$ &$0.32\pm0.02$ & $1.12\pm0.04$ & $0.96\pm0.01$ & $2.39\pm0.05$ \\ \cline{2-7}
 &  & $P_{c}^-(4312)$ &$0.47\pm0.02$ & $1.75\pm0.04$ & $1.46\pm0.02$ & $3.68\pm0.08$ \\
 & $J/\psi {\bar p}$ & $P_{c}^-(4440)$ &$0.98\pm0.03$ & $3.63\pm0.09$ & $2.99\pm0.06$ & $7.60\pm0.19$ \\
 &  & $P_{c}^-(4457)$ &$0.31\pm0.01$ & $1.11\pm0.02$ & $0.98\pm0.01$ & $2.39\pm0.03$ \\  \hline
 &  & $P_{c}^+(4312)$ & $0.65\pm0.01$ & $2.29\pm0.03$ & $1.93\pm0.02$ & $4.87\pm0.04$ \\
 & $J/\psi p$ & $P_{c}^+(4440)$ & $1.30\pm0.03$ & $4.80\pm0.06$ & $3.96\pm0.03$ & $10.05\pm0.07$ \\
13TeV &  & $P_{c}^+(4457)$ & $0.39\pm0.01$ & $1.48\pm0.01$ & $1.25\pm0.02$ & $3.11\pm0.01$ \\  \cline{2-7}
 &  & $P_{c}^-(4312)$ & $0.63\pm0.01$ & $2.30\pm0.03$ & $1.88\pm0.01$ & $4.82\pm0.05$ \\
 & $J/\psi {\bar p}$ & $P_{c}^-(4440)$ & $1.26\pm0.02$ & $4.73\pm0.05$ & $3.89\pm0.05$ & $9.88\pm0.03$ \\
 &  & $P_{c}^-(4457)$ & $0.38\pm0.01$ & $1.46\pm0.03$ & $1.21\pm0.02$ & $3.06\pm0.03$ \\  \hline \hline

\end{tabular} \label{paci1}
\end{table*}\end{large}

Fig.1 shows the yield of exotic state $P_{c}^{\pm}(4312)$, $P_{c}^{\pm}(4440)$ and $P_{c}^{\pm}(4457)$ for three structures varying with parameter $\Delta m$ from 0 to 20 MeV/c in $pp$ collisions at $\sqrt{s}=7$, 13~TeV. It can be seen that the yield distribution patterns of exotic state $P_{c}^{\pm}(4312)$, $P_{c}^{\pm}(4440)$ and $P_{c}^{\pm}(4457)$ for the three structures are all similar, which increases linearly as $\Delta m$ increases. Meanwhile, under the same $\Delta m$ and energy, the yields of exotic states with the nucleus-like structure are always larger than those with the molecular structure and pentaquark structure. Compared with $J/\psi p$ cluster and $J/\psi {\bar p}$ cluster, Fig.1 shows that $J/\psi {\bar p}$ is slightly less than $J/\psi p$. We can conclude that in the {\footnotesize PACIAE} model, it is easier to generate $P_{c}^+$ than antiparticle $P_{c}^-$. Furthermore, the yields of exotic states in $\sqrt{s}=13$~TeV is higher than those in $\sqrt{s}=7$~TeV, showing there exists an evident energy dependence.

\begin{figure*}[t]
\centering
\includegraphics[width=0.85\textwidth]{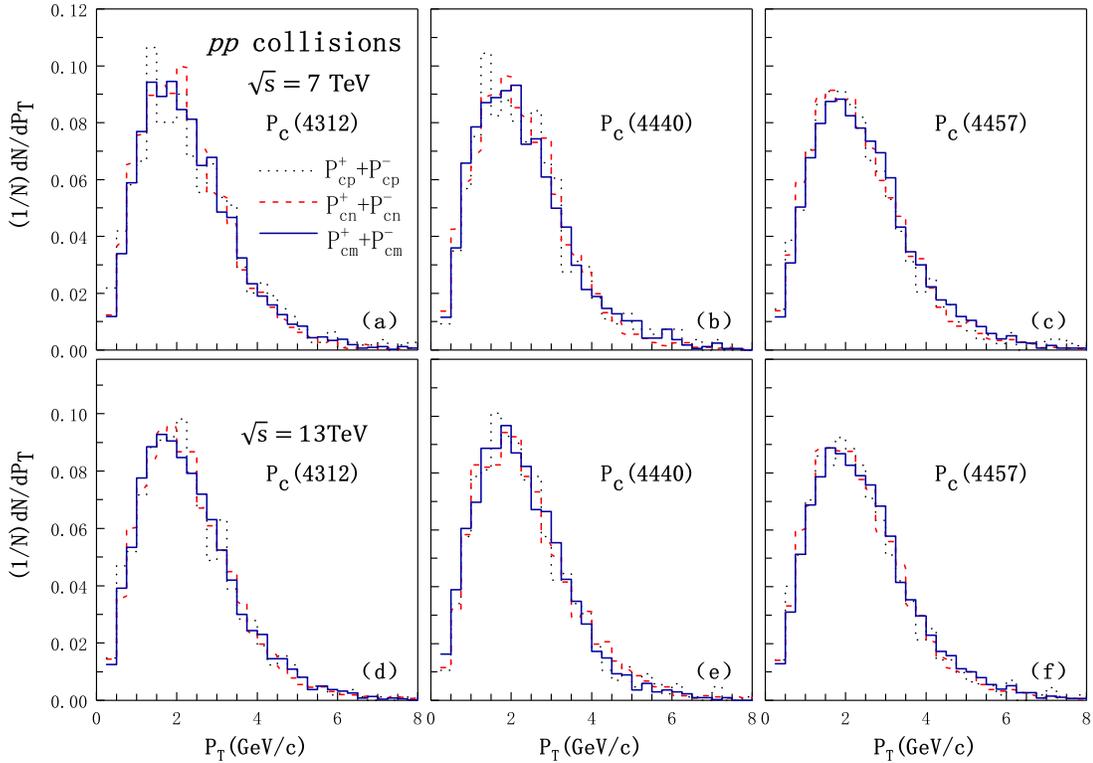}
\caption{The transverse momentum distribution of exotic states $P_{c}^{\pm}(4312)$, $P_{c}^{\pm}(4440)$ and $P_{c}^{\pm}(4457)$ with three different structures $(P_{cp}, P_{cn}, P_{cm})$ in $pp$ collisions at $\sqrt{s}=7$, 13~TeV. The data are computed by using PACIAE+DCPC model. Three different structures exotic states of $P_{cp}^{\pm}, P_{cn}^{\pm}, P_{cm}^{\pm}$ shown as black dotted, red dashed, and blue solid line, respectively. (a) $P_{c}^{\pm}(4312)$ at $\sqrt{s}=7$~TeV, (b) $P_{c}^{\pm}(4440)$ at $\sqrt{s}=7$~TeV, (c) $P_{c}^{\pm}(4457)$ at $\sqrt{s}=7$~TeV, (d) $P_{c}^{\pm}(4312)$ at $\sqrt{s}=13$~TeV, (e)$P_{c}^{\pm}(4440)$ at $\sqrt{s}=13$~TeV, (f) $P_{c}^{\pm}(4457)$ at $\sqrt{s}=13$~TeV. }
\label{tu3}
\end{figure*}

Note that, if the value of parameter $\Delta m$ is determined, we can predict the yields of exotic states $P_c$ from results in Fig.1. Here we take the parameter $\Delta m$ as the half-decay width of the $P_c \to J/\psi {p}$ mass spectrum, that is, $\Delta m=\Gamma/2$. The decay widths of the mass spectrum of $P_{c}^{\pm}(4312)$, $P_{c}^{\pm}(4440)$, and $P_{c}^{\pm}(4457)$ from LHCb~\cite{Ref8} are $9.8\pm{2.7}_{-4.5}^{+3.7}$, $20.6\pm{4.9}_{-10.1}^{+8.7}$ and $6.4\pm{2.0}_{-1.9}^{+5.7}$ MeV, respectively. Then their parameter values $\Delta m$ is taken as $\Delta m= 4.9$ MeV for $P_{c}^{\pm}(4312)$, as 10.3 MeV for $P_{c}^{\pm}(4440)$, and as 3.2 MeV for $P_{c}^{\pm}(4457)$. Thus we can predict the yields of $P_{c}^{\pm}(4312)$, $P_{c}^{\pm}(4440)$ and $P_{c}^{\pm}(4457)$ with three structures, as shown in Table 2. Overall, for the same exotic state $P_c$, the yield of molecular structure $P_{cm}$ is slightly less than that of nucleus-like structure $P_{cn}$, is about three times as much as that of pentaqaurk structure $P_{cp}$. Obviously, the yields of positive ($P_c^+$) and negative ($P_c^-$) exotic particles are the same within the error. In the last column of Table 2, the total yields of the three different structure of $P_{c}(4312)$, $P_{c}(4440)$ and $P_{c}(4457)$ particles produced in $pp$ collisions at $\sqrt{s}=7$, 13~TeV are shown, we could predict the total yield is on the order of magnitude $10^{-6}$.

\begin{figure*}[!htb]
\centering
\includegraphics[width=0.85\textwidth]{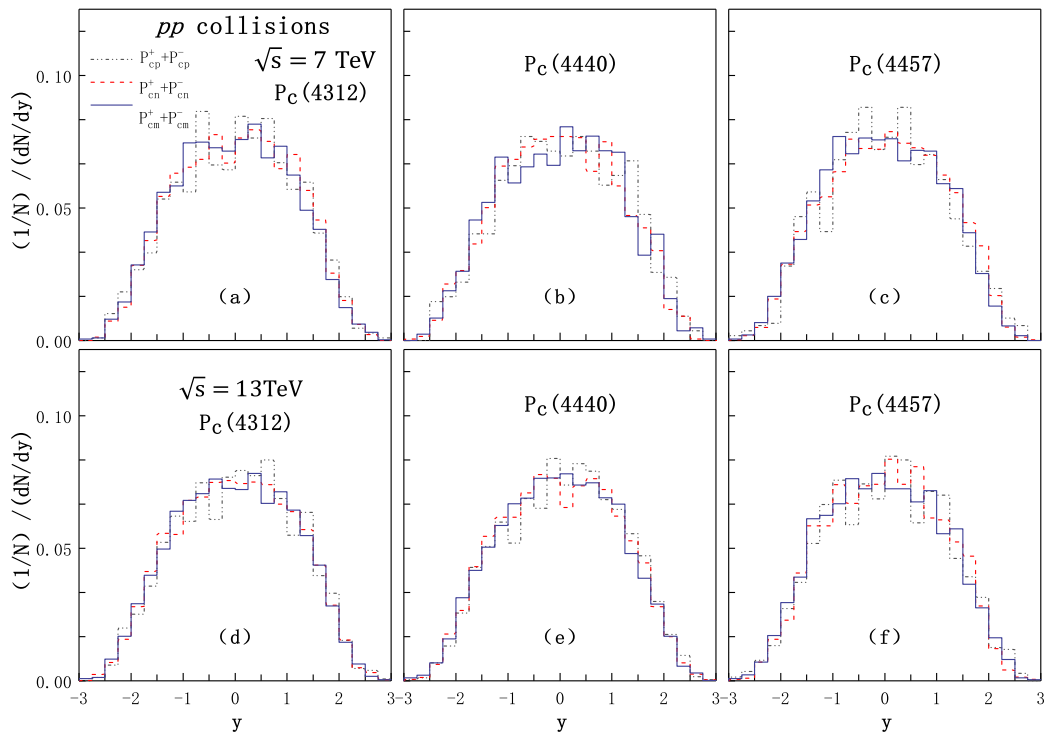}
\caption{ The rapidity distributions of exotic states $P_{c}^{\pm}(4312)$, $P_{c}^{\pm}(4440)$ and $P_{c}^{\pm}(4457)$ with three different structures $(P_{cp}, P_{cn}, P_{cm})$ in $pp$ collisions at $\sqrt{s}=7$, 13~TeV.}
\centering
\label{tu4}
\end{figure*}
Fig.2 shows the transverse momentum distribution of three different mass exotic states of $P_{c}^{\pm}(4312)$, $P_{c}^{\pm}(4440)$ and $P_{c}^{\pm}(4457)$ with three different structures $(P_{cp}, P_{cn}, P_{cm})$ in $pp$ collisions at $\sqrt{s}=7$, 13~TeV. The results are calculated by {\footnotesize PACIAE+DCPC} model. In each chart, the pentaquark structure, nucleus-like structure, and molecular structure are indicated by black dotted, red dashed, and blue solid line, respectively. From the Fig 2, it can be seen that the transverse momentum distribution of the exotic states of $P_{c}^{\pm}$ with three different mass of $P_{c}^{\pm}(4312)$, $P_{c}^{\pm}(4440)$, $P_{c}^{\pm}(4457)$, and three different structures of $P_{cp}^{\pm}$, $P_{cn}^{\pm}$, $P_{cm}^{\pm}$ are all generally similar. When the collision energy increases from 7 to 13 TeV, the peak value of the traverse momentum of different exotic states $P_{c}^{\pm}$ becomes slightly smaller, and the distribution becomes slightly wider.

The rapidity distributions of the exotic states $P_{c}^{\pm}(4312)$, $P_{c}^{\pm}(4440)$ and $P_{c}^{\pm}(4457)$ with three different structures are computed by {\footnotesize PACIAE+DCPC} model in $pp$ collisions at $\sqrt{s}=7$, 13~TeV, which are also presented in Fig.3. Obviously, one can see from Fig. 3 that there are symmetric distribution of rapidity distribution from -3 to 3. Here, rapidity distribution of the exotic states $P_{c}^{\pm}$ with the different mass, structures, and collisions energy are all similar.

\section{conclusions}

In this paper, we use the {\footnotesize PACIAE} model to simulate the production of final state particles with the $|y|<3$, $0<p_T<3$ GeV/c in $pp$ collisions at $\sqrt{s}=7$, 13~TeV. Then  by using {\footnotesize DCPC} model the exotic resonant state $P_{c}^{\pm}(4312)$, $P_{c}^{\pm}(4440)$, and $P_{c}^{\pm}(4457)$ are investigated based on the decaying $ P_c^\pm \to J/\psi p(\bar p)$. We propose that the exotic state $P_{c}^{\pm}(4312)$, $P_{c}^{\pm}(4440)$ and $P_{c}^{\pm}(4457)$ can be treated three possible structures of the pentaquark state ($P_{cp}^{\pm}$), the nucleus-like state ($P_{cn}^{\pm}$), and the molecular state ($P_{cm}^{\pm}$). Taking the half-life of the mass spectrum as the mass uncertainty parameter, $\Delta m=\Gamma/2$, the yield of exotic states $P_{c}^{\pm}(4312)$, $P_{c}^{\pm}(4440)$, and $P_{c}^{\pm}(4457)$ with three kinds of structure are predicted in $pp$ collisions at $\sqrt{s}=7$, 13~TeV, and the yield values are all on the order of $10^{-6}$. It is found that the yield of nucleus-like exotic states ($P_{cn}^{\pm}$) are greater than that of molecular states ($P_{cm}^{\pm}$), and the yield of molecular states ($P_{cm}^{\pm}$) are greater than that of pentaqaurk ($P_{cp}^{\pm}$). The yield of positive particles $P_{c}^+$ is slightly larger than that of antiparticles $P_{c}^-$, but within the error range they are approximately equal. Their yield increases with the increase of collision energy, i.e, the yield of the exotic states $P_{c}^\pm$ at $\sqrt{s}=7$~TeV is greater than that at $\sqrt{s}=13$~TeV. In addition, we study the transverse momentum distribution and the rapidity distributions of three exotic states $P_{c}^\pm$ with different structures. It is found that the transverse momentum distribution and the rapidity distributions of the exotic state $P_{c}^{\pm}$ with three different mass of $P_{c}^{\pm}(4312)$, $P_{c}^{\pm}(4440)$, $P_{c}^{\pm}(4457)$, three different structures of $P_{cp}^{\pm}$, $P_{cn}^{\pm}$, $P_{cm}^{\pm}$), and two different energies of 7, 13 TeV are all similar.

Finally, in order to find out further insight and understanding of the nature of the exotic resonant state $P_{c}^{\pm}(4312)$, $P_{c}^{\pm}(4440)$, and $P_{c}^{\pm}(4457)$ , we therefore suggest measurements of their production rates in $pp$ and heavy-ion collisions by the ALICE and LHCb experiments.

\textbf{Acknowledgments}

The work of Y. L. Xie is supported by the National Natural Science Foundation of China (12005196) and the Fundamental Research Funds for the Central Universities (G1323519234), of D. M. Zhou is supported by the NSFC (11705167), and of G. Chen is supported by the National Natural Science Foundation of China (NSFC) (11475149).

\textbf{Data Availability Statement} This manuscript has no associated data
or the data will not be deposited. [Authors¡¯ comment: The data used to
support the findings of this study are available from the corresponding
author upon request.]

\end{document}